%% file: eprint.tex
\documentclass[10pt, paper=a4, UKenglish]{article}

\usepackage{amsmath}
\usepackage{xspace}
\usepackage{graphicx}
\usepackage{textcomp,gensymb}
\graphicspath {{./images/}}

\usepackage{hyperref}
\hypersetup{
     colorlinks=true,       
     linkcolor=red,         
     citecolor=magenta,     
     filecolor=magenta,     
     urlcolor=blue          
}

\newcommand{\pt}{\ensuremath{p_\text{T}}\xspace}
\newcommand{\xju}[1]{\textcolor{red}{#1}}

\def\Title#1{\begin{center} {\Large #1 } \end{center}}
\def\Author#1{\begin{center}{ \sc #1} \end{center}}
\def\Address#1{\begin{center}{ \it #1} \end{center}}

\newcommand\pubblock{\rightline{\begin{tabular}{l} Proceedings of the CTD 2022\\ \pubnumber\\
         \pubdate  \end{tabular}}}

\newenvironment{Abstract}{\begin{quotation} \begin{center} 
             \large ABSTRACT \end{center}\bigskip 
      \begin{center}\begin{large}}{\end{large}\end{center} \end{quotation}}

\newenvironment{Presented}{\begin{quotation} \begin{center} 
             PRESENTED AT\end{center}\bigskip 
      \begin{center}\begin{large}}{\end{large}\end{center} \end{quotation}}

\def\Acknowledgements{\bigskip  \bigskip \begin{center} \begin{large}
      \bf ACKNOWLEDGEMENTS \end{large}\end{center}}

\input econfmacros.tex

\usepackage{color}
\usepackage{lineno}
\usepackage{subfig}
\usepackage{hyperref}
\usepackage{comment}
\usepackage{lipsum}

\newcommand\pubnumber{PROC-CTD2022-23}

\newcommand\pubdate{\today}

\def\affiliation{
On behalf of PANDA Collaboration, \\
Department of Physics and Astronomy, Uppsala University \\
Lägerhyddsvägen 1, 75237 Uppsala, Sweden}

\newcommand{\conference}{Connecting the Dots Workshop (CTD 2022)\\
May 31 - June 2, 2022}

\usepackage{fancyhdr}
\definecolor{mygrey}{RGB}{105,105,105}
\begin{document}


\large
\begin{titlepage}
\pubblock

\vfill
\Title{Track Reconstruction using Geometric Deep Learning in the Straw Tube Tracker (STT) at the PANDA Experiment}
\vfill

\Author{Adeel Akram}
\Address{\affiliation}

\Author{Xiangyang Ju}
\Address{Lawrence Berkeley National Laboratory, \\ Berkeley, CA 94720, United States}
\vfill

\begin{Abstract}
The PANDA (anti-Proton ANnihilation at DArmstadt) experiment at the Facility for Anti-proton and Ion Research will study strong interactions at the scale at which quarks are confined to form hadrons. A continuous beam of anti-protons, provided by the High Energy Storage Ring (HESR), will impinge on a fixed hydrogen target. The anti-proton beam momentum spanning from 1.5~GeV\footnote{Natural units, c=1} to 15~GeV~\cite{physics2009report} will create optimal conditions for studying many different aspects of hadron physics, including hyperon physics.

Precision physics studies require a highly efficient particle track reconstruction. The Straw Tube Tracker in PANDA is the main component for that purpose.  It has a hexagonal geometry, consisting of 4224 gas-filled tubes arranged in 26 layers and six sectors. However, the challenge is reconstructing low-momentum charged particles given the complex detector geometry and the strongly curved particle trajectory. This paper presents the first application of a geometric deep learning (GDL) pipeline, inspired by the Exa.TrkX project, to the track reconstruction task in the PANDA experiment. The pipeline is suitable for graph-based data and uses graph neural networks (GNNs) as the underlying learning algorithm. The pipeline reconstructs more than 95\% of particle tracks and creates less than 0.3\% of fake tracks. The promising results make the pipeline a strong candidate algorithm for the experiment.

\end{Abstract}

\vfill

\begin{Presented}
\conference
\end{Presented}
\vfill
\end{titlepage}
\def\thefootnote{\fnsymbol{footnote}}
\setcounter{footnote}{0}
%

\normalsize 


\section{Introduction} \label{intro}

The PANDA experiment~\cite{physics2009report} will provide unique opportunities to study Quantum Chromodynamics (QCD) in the confinement domain where an anti-proton beam impinges on a stationary target (proton or heavy atoms). Dynamic beam momentum, ranging from 1.5~GeV to 15 GeV, allows in-depth studies of Nucleon Structure, Strangeness Physics, Charm and Exotics and Hadrons in Nuclei~\cite{phase1}. Generally, these studies are performed using the PandaRoot~\cite{Spataro2011} simulation framework to simulate nuclear reactions, for example, the $\Lambda$ hyperon production in $\bar{p}p \rightarrow \bar{\Lambda} \Lambda \rightarrow \bar{p}\pi^+ + p \pi^-$ reaction. To successfully reconstruct a physics reaction, one needs high-precision track reconstruction. The main challenge lies in reconstructing particles with low transversal momentum (\pt), for example, pions. The weekly decaying hyperons, like $\Lambda$ particles, fly a considerable distance ($c\tau (\Lambda) = 7.89 \textrm{ cm}$) in the detector before decaying. Thus, the presence of displaced secondary vertices is an additional challenge. Tracking algorithms explored in Ref.~\cite{aalikce} try different combinations of detector measurements using a helix fitting to determine the quality of track candidates. However, our work presents the first application of a GNN as an alternative approach to track reconstruction in PANDA. The GNNs are better suited for non-Euclidean data because of the expressive power of the graph structure.

The paper is organized as follows. A brief introduction to the PANDA experiment is given in Section~\ref{sec:panda}. The GDL pipeline is discussed in Section~\ref{sec:pipeline} whereas results, \textit{track evaluation}, are given in Section~\ref{sec:track_evaluation}. Finally, the conclusion is given in Section~\ref{sec:conclusion}.

\section{The PANDA Experiment at FAIR} \label{sec:panda}

\begin{figure}[!htb]
  \centering
  \includegraphics[width=0.9\linewidth]{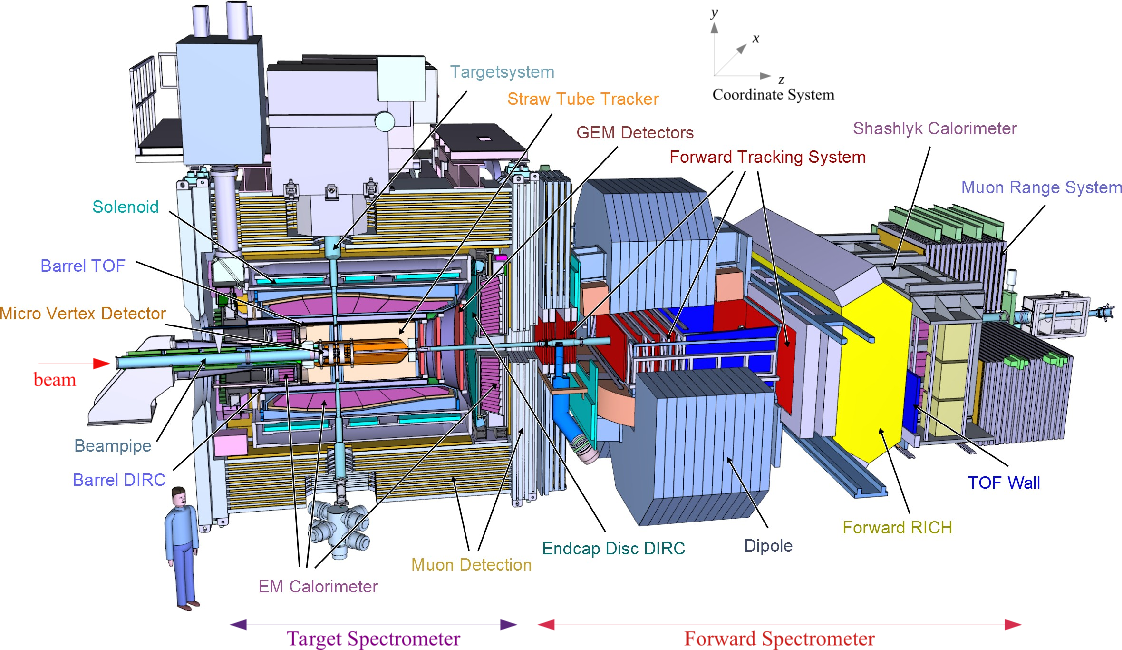}
  \caption{Schematic of the PANDA Experiment}
  \label{fig:panda}
\end{figure}

The PANDA (anti-Proton ANnihilation at DArmstadt) experiment is currently under construction at FAIR (Facility for Anti-proton and Ion Research)~\cite{Gutbrod54062, Gutbrod54068}. Figure~\ref{fig:panda} shows a schematic diagram of the PANDA experiment. The detector provides almost $4\pi$ coverage surrounding the interaction point (IP) and will record data at the expected interaction rate of $20 \textrm{ MHz}$. The Straw Tube Tracker (STT) in the target spectrometer is the most demanding detector for reconstructing particle trajectories.

\begin{figure}[!htb]
  \centering
  \includegraphics[width=0.7\linewidth]{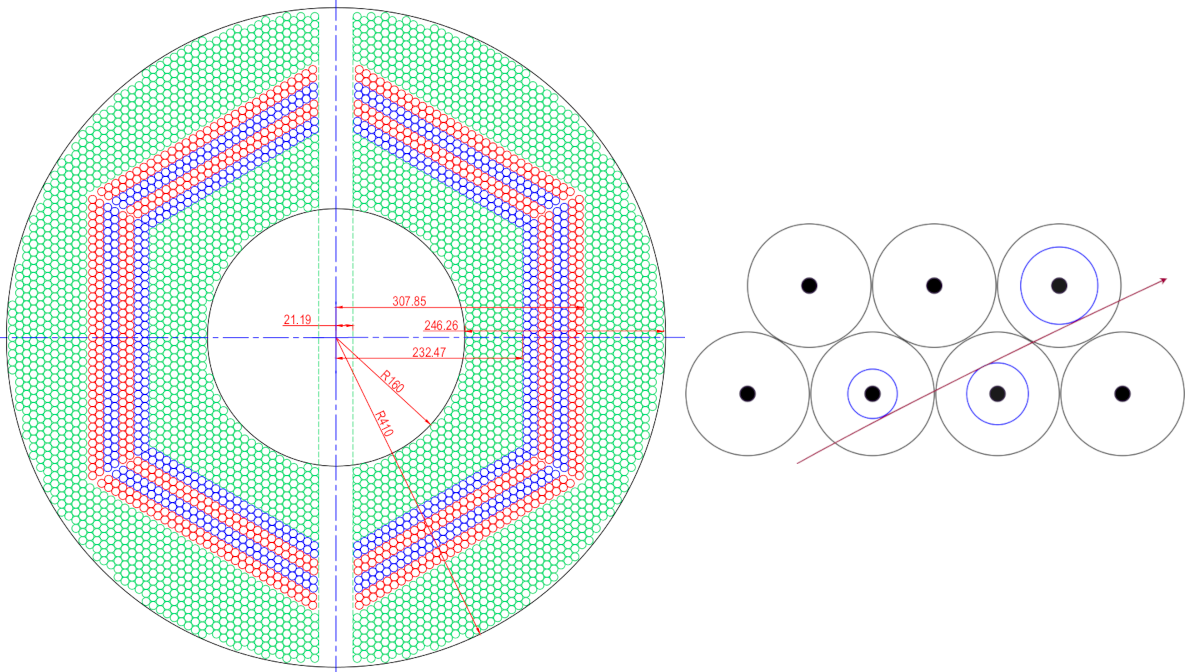}
  \caption{Left: a cross-sectional view of the STT detector. Right: zoom-in view of strew tubes in black circles and the isochrone radius in blue circles and a track in red.}
  \label{fig:stt}
\end{figure}

The STT consists of 4224 straw tubes, the single-channel drift tubes, distributed in 27 layers and six sectors in a hexagonal shape, as shown in Figure~\ref{fig:stt} (left). The green tubes (15 - 19 layers) are parallel to the beam axis intended for $xy-$position whereas blue and red tubes (8 layers) are skewed to $\pm 2.9^{\degree}$ polar angle with respect to the beam axis for reconstructing the $z-$ component. The STT detector angular acceptance in polar angle, $\theta$, is from $22^{\degree}$ to $140^{\degree}$. A particle needs a minimum \pt of 50~MeV to arrive at the innermost layer and a minimum \pt of 100~MeV to traverse through the STT fully\footnote{Under solenoid magnetic field of 2 T}.

When a charged particle traverses the STT detector, it ionizes the gas inside the tube. The ionized electrons then drift towards the centre of the tube, where an anode wire creates a \textit{hit}. The $xy$ position of the hit is the position of the anode wire. The distance between the closest approach of the particle trajectory to the anode wire is the \textit{isochrone radius}. We use both the $xy$ position and the isochrone radius as inputs. It is possible to reconstruct the $z$-component of the hit position by using the skewed straw layers, but that would require a dedicated algorithm to achieve high accuracy. We leave that work to the future.

\section{Geometric Deep Learning Pipeline}\label{sec:pipeline}
\begin{figure}[!htb]
  \centering
  \includegraphics[width=0.85\linewidth]{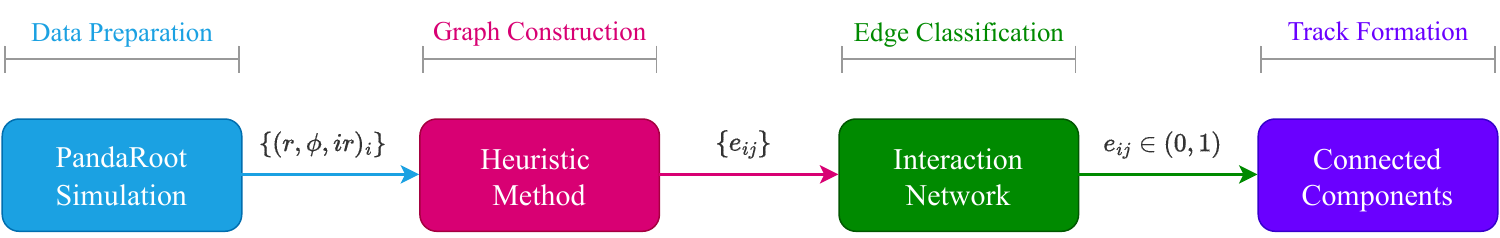}
  \caption{Stages of Geometric Deep Learning pipeline used for reconstructing tracks with the STT detector.}
  \label{fig:pipeline}
\end{figure}

To perform track reconstruction using \textit{Geometric Deep Learning (GDL)}, we build a complete pipeline on top of the Exa.TrkX \cite{ExaTrkX} pipeline. The core idea is to first represent a collision event as a graph in which nodes are the hits recorded by the STT detector and edges are connections between two adjacent hits,  and then to train a Graph Neural Network (GNN) to learn the relational information between hits. Our pipeline consists of four discrete stages (see Figure~\ref{fig:pipeline}): data preparation, graph construction, edge classification, and track formation.

\subsection{Data Preparation}

The input data is simulated using the PandaRoot analysis framework~\cite{Spataro2011}. Five $\mu^+\mu^-$ pairs per event are generated using the particle gun in the momentum range of 100~MeV/c to 1.5~GeV/c. Muons are isotropically generated within the geometric acceptance of the STT detector (see Section~\ref{sec:panda}). The choice of this relatively low momentum range is motivated by the fact that the existing algorithms have difficulties in reconstructing low-\pt particle trajectories because these trajectories are  spiralling, strongly curved, and overlapping~\cite{aalikce, Regina1620060}.

A total of $100,000$ events are generated, containing about a million tracks. The data is split into $90 \%$ for training, $5 \%$ for validation, and $5\%$ for testing.

\subsection{Graph Construction}

\begin{figure}[!htb]
  \centering
  \subfloat{\includegraphics[width=0.47\linewidth]{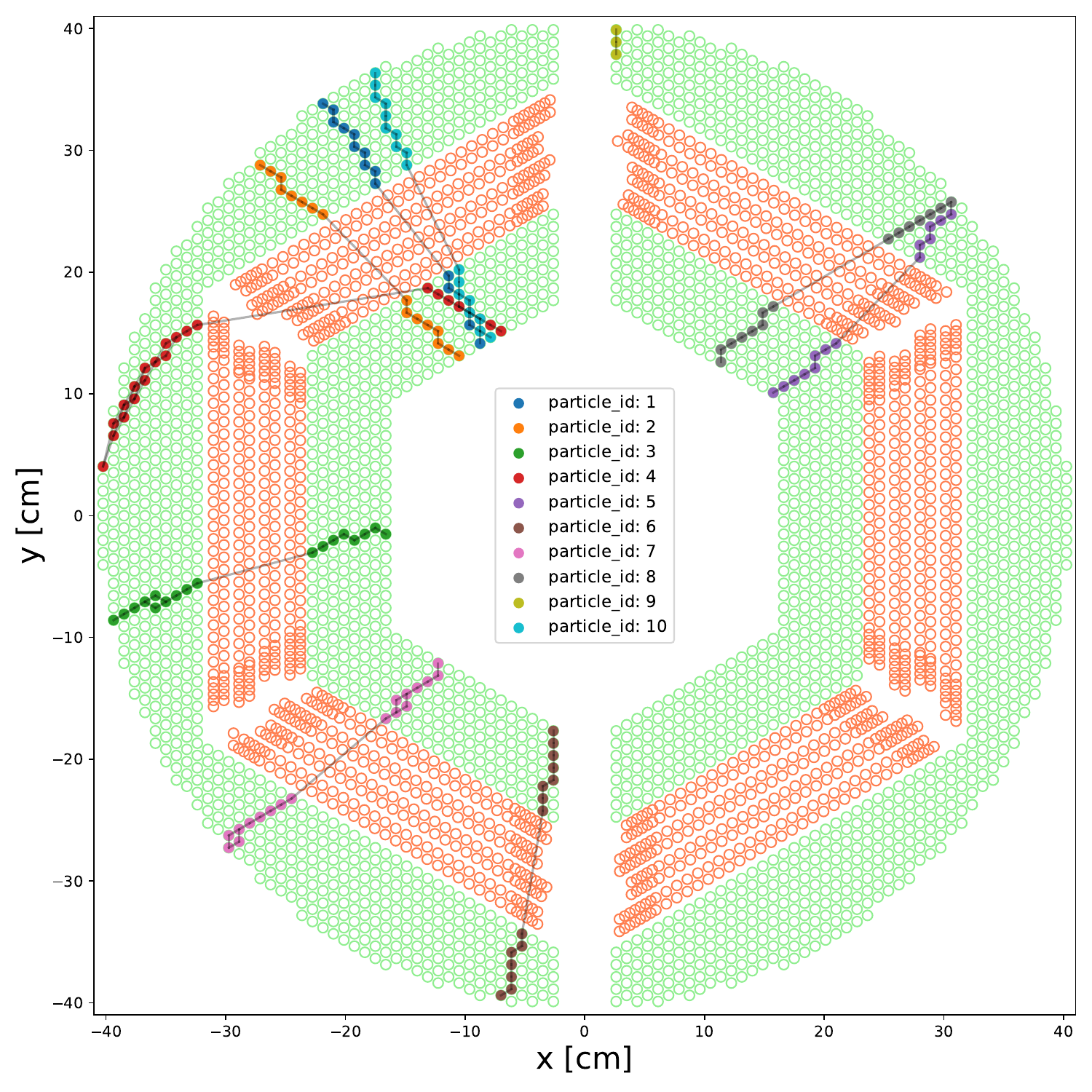}}
  \quad
  \subfloat{\includegraphics[width=0.47\linewidth]{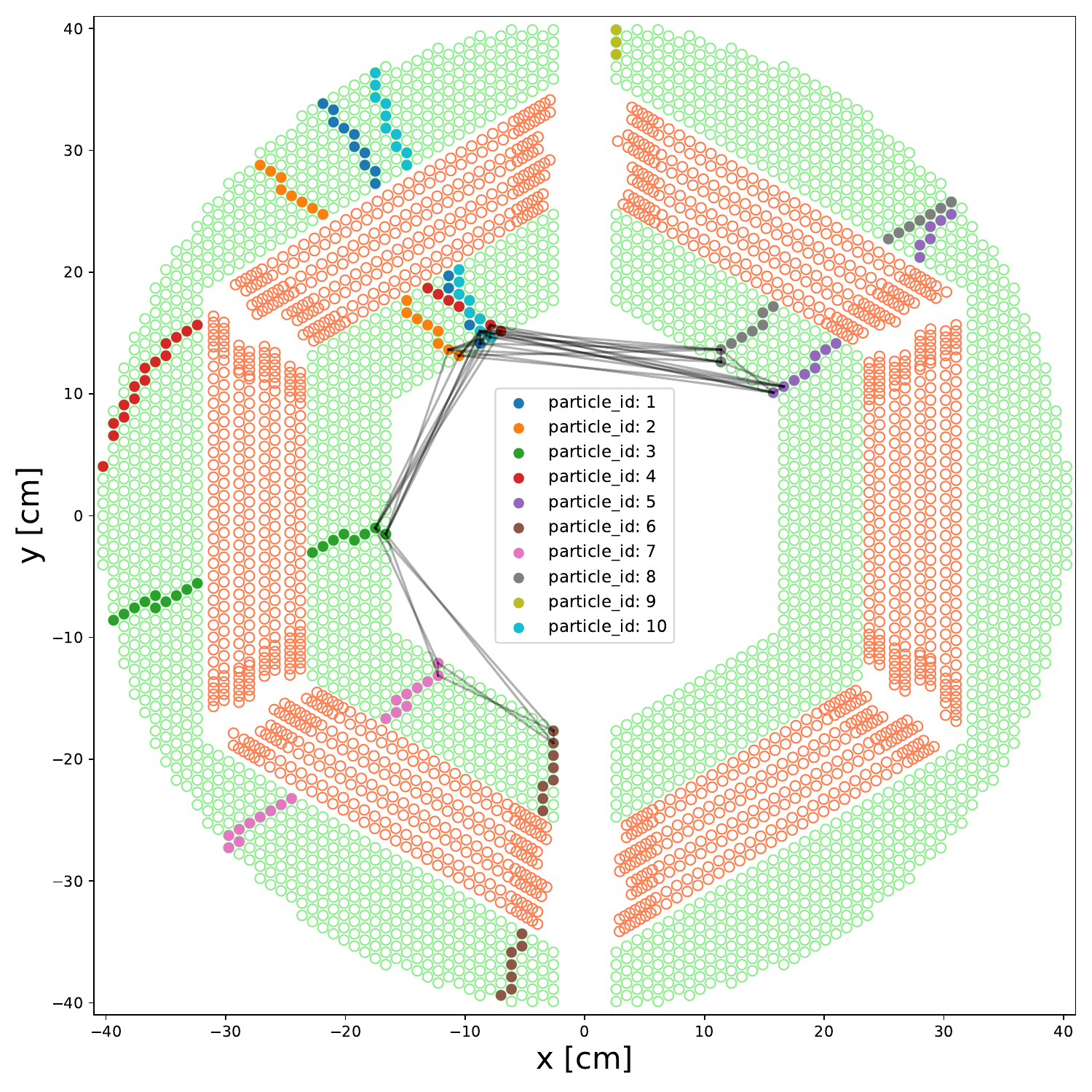}}
  \caption{Graph representation of an event: \emph{(left)} True Graph, \emph{(right)} Input Graph} 
  \label{fig:example-event}
\end{figure}

The graph construction step is to convert the raw input data, represented as point clouds, into bi-directed  graphs. To that end, we use a heuristic method that captures all true edges while it keeps fake connections low. The method builds all combinatorial connections between hits in two consecutive radially outward layers in nearby sectors. Figure~\ref{fig:example-event} shows the true graph (left) where edges are connections between hits coming from the same particle and the input graph (right) in which only the connections between the innermost two layers are shown. If a particle leaves two hits in the same layer, both hits are included in the graph.

\subsection{Edge Classification}

\begin{figure}[!htb]
  \centering
  \includegraphics[width=0.8\linewidth]{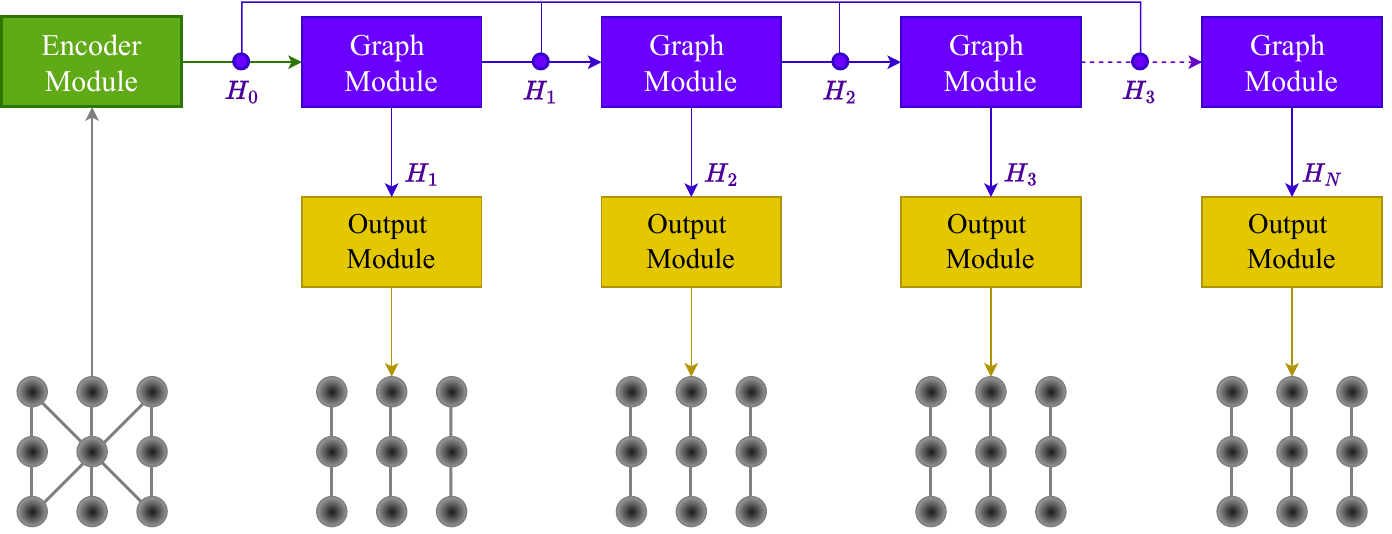}
  \caption{Interaction Graph Neural Network (Interaction GNN)}
  \label{fig:interactionGNN}
\end{figure}

We use the same Graph Neural Network architecture used in Ref.~\cite{Ju:2021ayy}, also known as Interaction GNN. The schematic of this network is shown in Figure~\ref{fig:interactionGNN}.

The Interaction GNN contains three sub-networks: the Encoder Module, the Graph Module and the Output Module. The Encoder Module comprises a node network and an edge network, encoding input node features into a vector of hidden features and creating edge features with the neighbouring node features. The Graph Module is an interaction network~\cite{2018arXiv180601261B} in which aggregated neighbouring edge features are passed to the node network, and the neighbouring node features are passed to the edge network. This way, information is exchanged between nodes and edges, namely message passing. The message passing step is performed $N=8$ times. The output from the Graph Module is passed through the output module that performs binary classification of edges using the cross-entropy loss. As a result, an edge score is obtained for each edge.

\begin{figure}[!htb]
  \centering
  \subfloat{\includegraphics[width=0.47\linewidth]{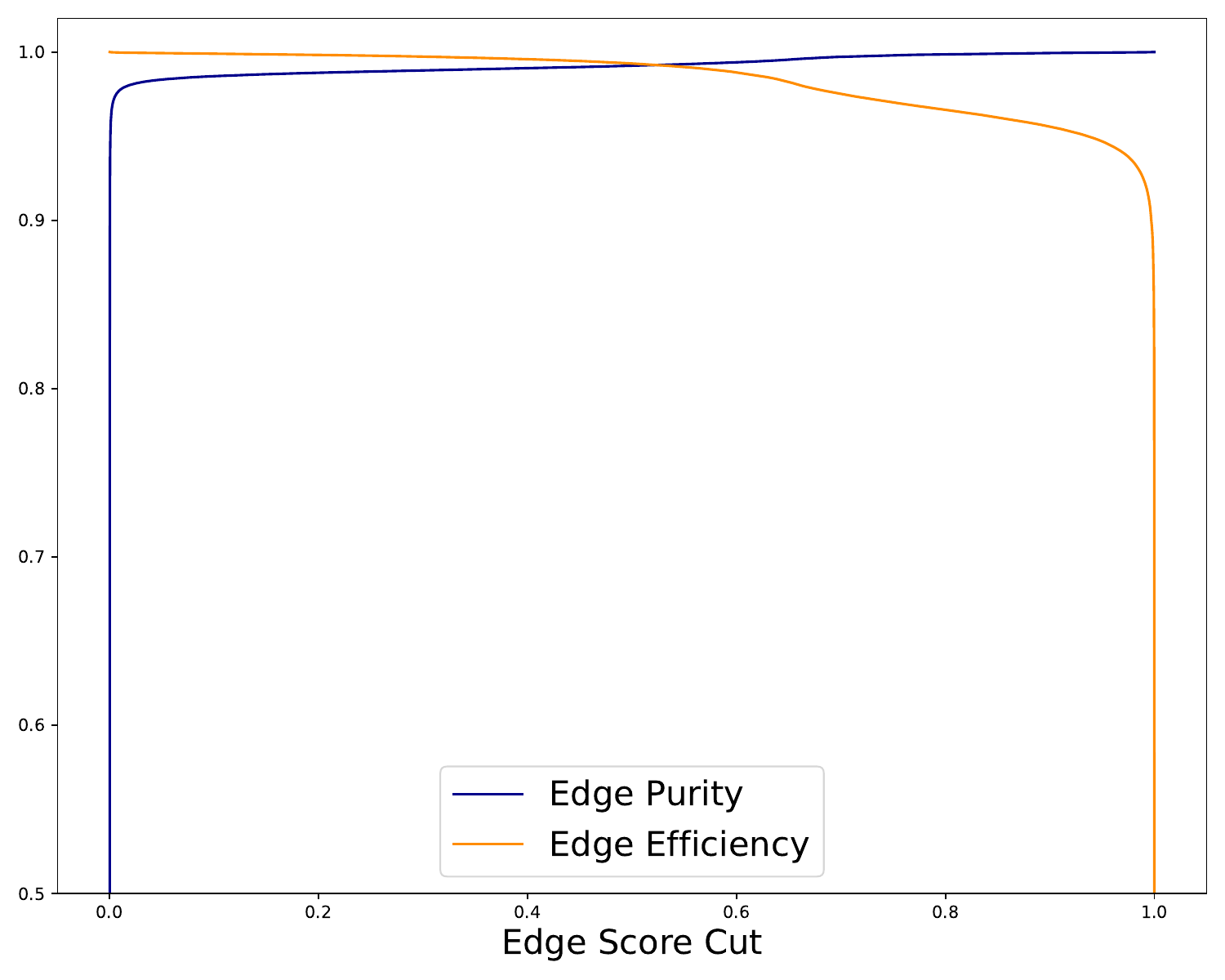}}
  \quad
  \subfloat{\includegraphics[width=0.47\linewidth]{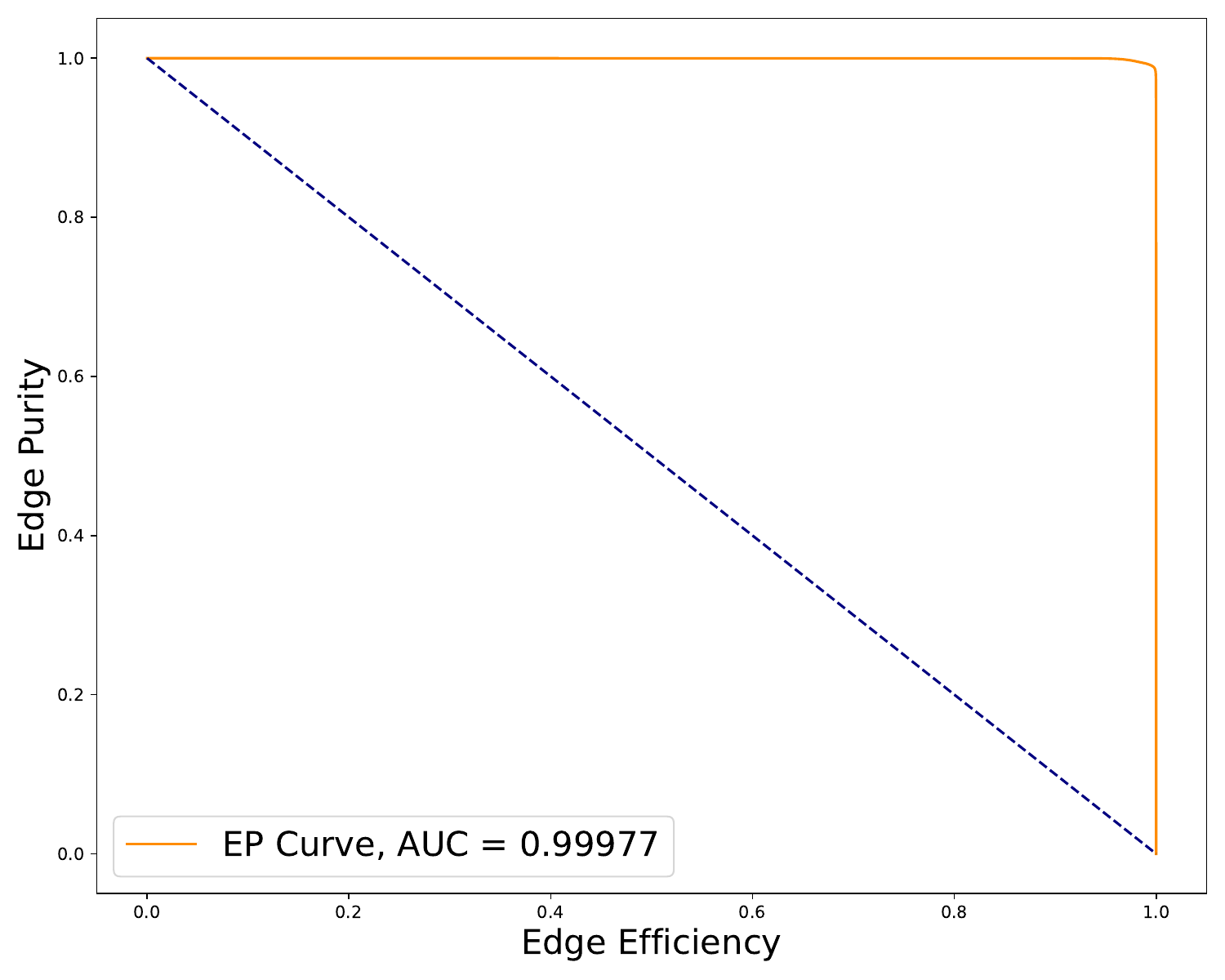}}
  \caption{Model evaluation: $\epsilon_{E}$ and $p_{E}$ as a function of edge score cut \emph{(left)}, ROC with $\textrm{AUC} > 0.9998$ \emph{(right)}}
  \label{fig:epc_curve}
\end{figure}

The performance of the interaction GNN is evaluated using the Area Under the Curve (AUC) of the Receiver Operating Characteristic (ROC) curve. The higher the AUC value, the better the model performance. In addition, we define the edge efficiency ($\epsilon_E$) as the fraction of true edges being selected and the edge purity ($p_E$) as the fraction of true edges in the selected edges. Figure~\ref{fig:epc_curve}(a) shows the edge efficiency and edge purity distribution as a function of edge score cuts and Figure~\ref{fig:epc_curve}(b) shows the ROC curve built from $\epsilon_E$ and $p_E$ as defined above. The AUC for the Interaction GNN is $0.9998$.

\subsection{Track Formation}\label{sec:track_building}

Track formation is to create a list of hits, also known as \emph{track candidates}, from the edge scores predicted by the interaction GNN. To that end, we use the \emph{connected component} algorithm in the Graph theory~\cite{wcc}. We first remove edges with scores less than 0.25 and then find \emph{weakly connected components} in the graph. Each found component is a track candidate. 

\section{Results}\label{sec:track_evaluation}

We define the Track Efficiency ($\epsilon_{\textrm{phys}}$, $\epsilon_{\textrm{tech.}}$), Track Purity, and Fake Rate to evaluate the performance of the track finding pipeline. These metrics are defined as follows:

\begin{equation}\label{eq:epc}
\begin{aligned}
    \epsilon_{\textrm{phys}} &= \frac{N_{particles} (\textrm{selected, matched})}{N_{particles} (\textrm{selected})} \\
    \epsilon_{\textrm{tech.}} &= \frac{N_{particles} (\textrm{selected, reconstructable, matched})}{N_{particles} (\textrm{selected, reconstructable})} \\
    \textrm{Purity} &= \frac{N_{\text{reco\_tracks}} (\textrm{selected, matched})}{N_{\text{reco\_tracks}} (\textrm{selected})} \equiv 1 - \textrm{Fake Rate}
\end{aligned}
\end{equation}

where $N_{particles} (\textrm{selected})$ is the number of particles generated within the STT acceptance, and $N_{particles}$ (selected, reconstructable) is the number of generated  particles that leave at least \xju{4} hits in the detector, namely the reconstructable generated particles. Generated particles are reconstructed if they can be matched~\footnote{the matching criteria is that the number of hits shared by the particle track and reconstructed track is larger than 50\% of the number of hits of the particle track and larger than 50\% of the number of hits of the reconstructed track. A particle track is formed by a list of hits created by the particle.} to at least one reconstructed track. Similarly, $N_{\text{reco\_tracks}}(\text{selected})$ is the number of reconstructed tracks containing at least 4 hits. The fake rate is defined as the fraction of reconstructed tracks not matching any particle tracks, and the duplication rate is defined as the rate at which a particle track is matched to more than one reconstructed track.

\begin{figure}[!htb]
  \centering
  \subfloat{\includegraphics[width=0.47\linewidth]{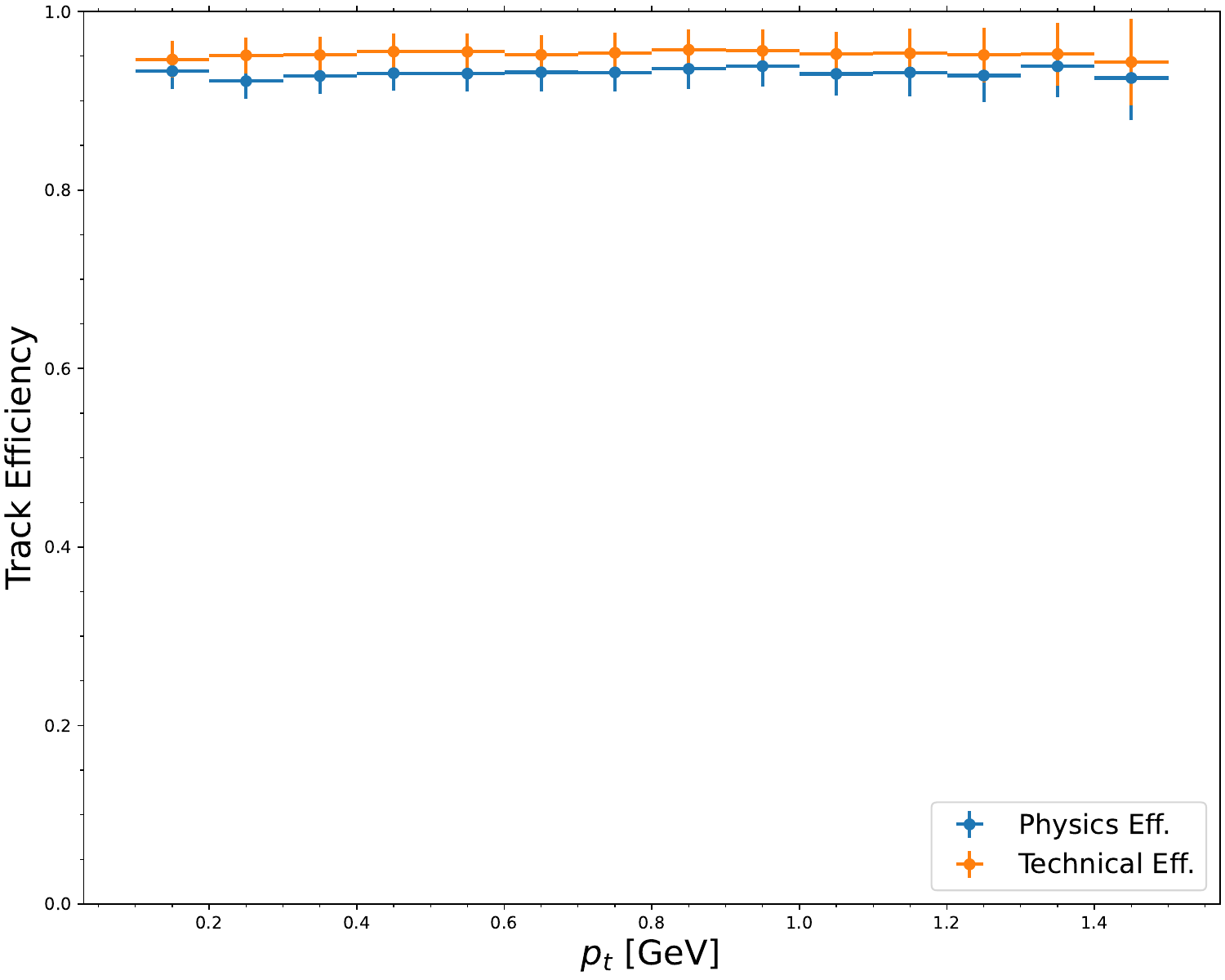}}
  \quad
  \subfloat{\includegraphics[width=0.47\linewidth]{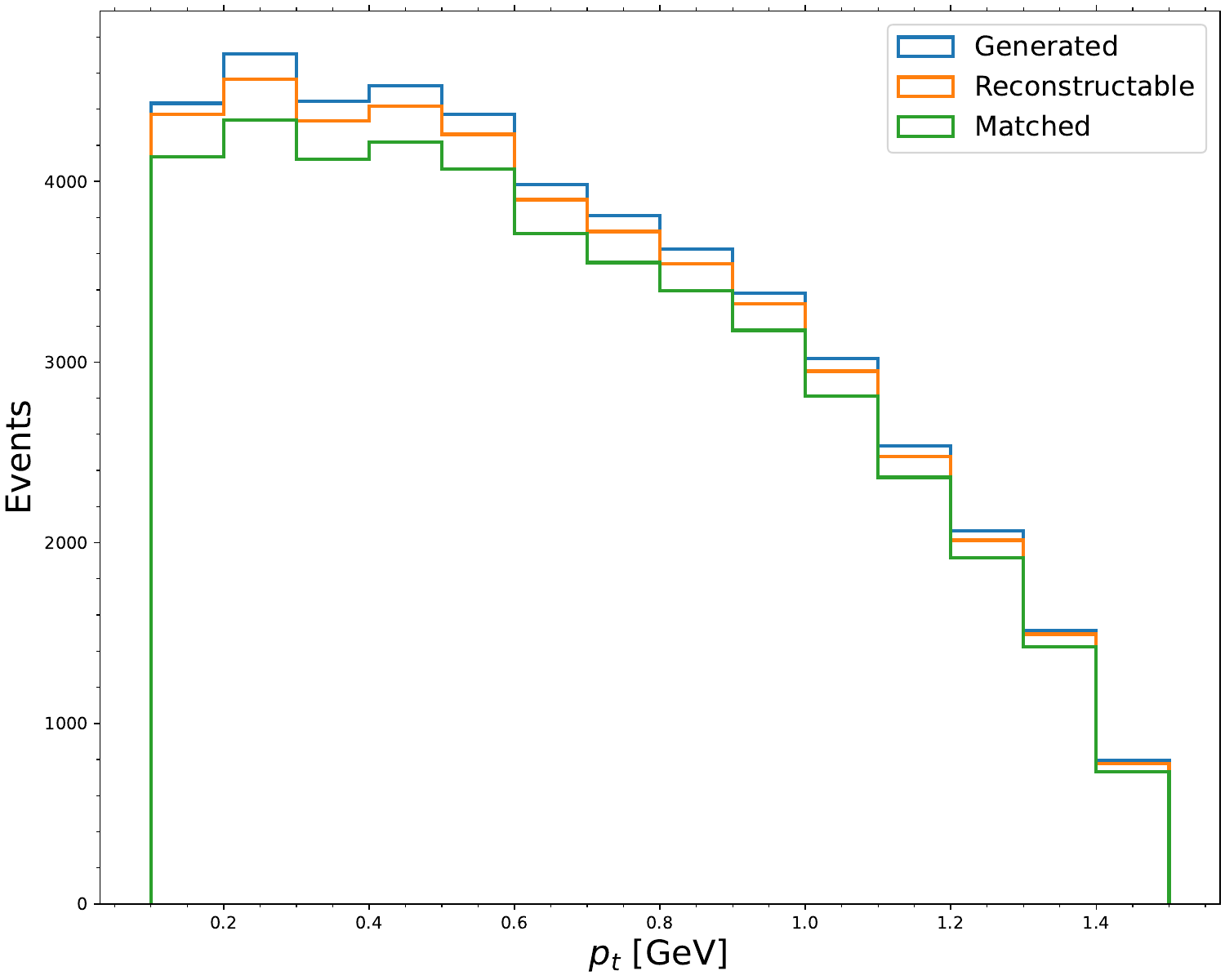}}
  \caption{Track evaluation: \emph{(left)} Track efficiency \emph{vs.} \pt, \emph{(right)} Number of generated, reconstructable and matched tracks at various \pt values.}
  \label{fig:tracking_eval}
\end{figure}

A test dataset with 5000 events is used for the evaluation. The overall $\epsilon_{\textrm{phys}}$ is 94.6\% with a fake rate of 0.423\% and a duplication rate of 16.6\%. Figure~\ref{fig:tracking_eval}(a) shows the $\epsilon_{\textrm{phys}}$ and $\epsilon_{\textrm{tech.}}$ over different \pt values. It is expected that the $\epsilon_{\textrm{tech.}}$ will be slightly greater than the $\epsilon_{\textrm{phys}}$ due to the inefficiency of the detector. Figure~\ref{fig:tracking_eval}(b) shows the number of generated, reconstructable and matched events at various \pt values.

The inefficiency and the high duplication rate are caused by crossing or overlapping tracks. For events consisting of well-separated tracks in an event, the $\epsilon_{\textrm{phys}}$ is  $97.6\%$ with a fake rate of 0.220\% and a duplication rate of 7.92\%. The remaining efficiency loss, in some cases, is caused by extremely low \pt tracks where a single track is broken into multiple sub-tracks, leading to low efficiency and a high duplication rate. Further investigation reveals that our track labelling method fails to assign shared hits to different tracks properly.


\section{Conclusions}\label{sec:conclusion}
The PANDA experiment will be a unique facility to study strong interactions. We presented a GDL-based track reconstruction pipeline for the PANDA experiment. The pipeline  achieves high reconstruction efficiencies at a low rate of fake tracks. However, the rate of duplicate tracks is high due to shared hits between tracks. A more robust track labelling algorithm will be investigated.

In the future, we would like to compare the GNN-based pipeline with other track reconstruction algorithms currently under development at PANDA and test the pipeline with different physics processes. We will also test the robustness of the pipeline and add the track parameter estimation.


\Acknowledgements
This work has been made possible through the Knut and Alice Wallenberg Foundation (Sweden) under contract no. 2016.0157 (PI: K. Schönning). In addition, this research used resources of the National Energy Research Scientific Computing Center (NERSC), a U.S. Department of Energy Office of Science User Facility located at Lawrence Berkeley National Laboratory, operated under Contract No. DE-AC02-05CH11231 using NERSC award ERCAP-0021226.


\bibliographystyle{JHEP}
\bibliography{eprint.bib}

\end{document}

%% file: econfmacros.tex
\def\beq{\begin{equation}}
\def\eeq#1{\label{#1}\end{equation}}
\def\eeqn{\end{equation}}

\def\beqa{\begin{eqnarray}}
\def\eeqa#1{\label{#1}\end{eqnarray}}
\def\eeqan{\end{eqnarray}}

\let\bar=\overbar

\def\Dslash{\not{\hbox{\kern-4pt $D$}}}
\def\dslash{\not{\hbox{\kern-2pt $\del$}}}

\def\msb{{\bar{\ssstyle M \kern -1pt S}}}

